\documentclass[twocolumn,groupedaddress,showpacs,preprintnumbers,amsmath,amssymb]{revtex4}

\usepackage{graphicx}
\usepackage{dcolumn}
\usepackage{bm}

\begin{document}
\preprint{Submitted to Journal of Chemical Physics}

\title{Vibron-polaron in $\alpha$-helices. I. Single-vibron states}

\author{Cyril Falvo}
\author{Vincent Pouthier}
\email{vincent.pouthier@univ-fcomte.fr}
\affiliation{Laboratoire de Physique Mol\'{e}culaire, UMR CNRS 6624. Facult\'{e} des
Sciences - La Bouloie, \\ Universit\'{e} de Franche-Comt\'{e}, 25030 Besan\c {c}on cedex, 
France.}

\date{\today}

\begin{abstract}

The vibron dynamics associated to amide-I vibrations in a 3D $\alpha$-helix is described according to a generalized Davydov model. The helix is modeled by three spines of hydrogen-bonded peptide units linked via covalent bonds. To remove the intramolecular anharmonicity of each amide-I mode and to renormalized the vibron-phonon coupling, two unitary transformation have been applied to reach the dressed anharmonic vibron point of view. It is shown that the vibron dynamics results from the competition between inter-spine and intra-spine vibron hops and that the two kinds of hopping processes do not experience the same dressing mechanism. 
Therefore, at low temperature (or weak vibron-phonon coupling), the polaron behaves as an undressed vibron delocalized over all the spines whereas at biological temperature (or strong vibron-phonon coupling), the dressing effect strongly reduces the vibrational exchanges between different spines. As a result the polaron propagates along a single spine as in the 1D Davydov model. Although the helix supports both acoustical and optical phonons, this feature originates in the coupling between the vibron and the acoustical phonons, only. Finally, the lattice distortion which accompanies the polaron has been determined and it is shown that residues located on the excited spine are subjected to a stronger deformation than the other residues.    

\end{abstract}


\maketitle

\section{Introduction}

In living systems, the energy released by the hydrolysis of adenosine triphophate (ATP) is a universal energy source allowing many biological processes such as muscle contraction, active transport or enzyme catalysis. However, the fundamental question arises whether this energy can be transported from the active sites of the living cell to other regions without being dispersed or dissipated. 

This question was first pointed out by Davydov and co-workers in the 70th to explain the energy transport in $\alpha$-helices \cite{kn:davydov,kn:davydov1}. The main idea is that the released energy, partially stored in the high-frequency amide-I vibration of a peptide group, can be transported from one end of the helix to the other. The dipole-dipole coupling between the different peptide groups leads to the delocalization of the internal vibrations and to the formation of vibrons. However, the interaction between the vibrons and the phonons of the helix induces a nonlinear dynamics which counterbalances the dispersion and yields the creation of the so-called Davydov soliton (for a recent review, see for instance Refs. \cite{kn:scott1,kn:chris}).

Unfortunately, no clear evidence has yet been found for the existence of solitons in real proteins and it has been suggested by Brown  and co-workers \cite{kn:brown1,kn:brown2} and by Ivic and co-workers \cite{kn:ivic1,kn:ivic2,kn:ivic3} that the solution is rather a small polaron than a soliton. Indeed, this problem exhibits two asymptotic solutions depending on the values taken by three relevant parameters, i.e. the vibron bandwidth, the phonon cutoff frequency and the strength of the vibron phonon coupling. When the vibron bandwidth is greater than the phonon cutoff frequency, the adiabatic limit is reached so that the phonons behave in a classical way and create a quasistatic potential well responsible for the trapping of the vibron according to the Davydov soliton. By contrast, when the vibron bandwidth is lesser than the phonon cutoff  frequency, the situation corresponds to the non-adiabatic limit in which the quantum behavior of the phonons plays a crucial role. A vibron is thus dressed by a virtual cloud of phonons which yields a lattice distortion essentially located on a single site and which follows instantaneously the vibron. The vibron dressed by the lattice distortion forms the small polaron and, as mentioned by the authors, this situation corresponds to protein dynamics. Recently, the polaron approach has been improved to characterize the two-polaron energy spectrum with a special emphasis onto the interplay between the dressing effect and the intramolecular anharmonicity of each amide-I vibration \cite{kn:pouthier1,kn:pouthier2}. Both features favor the formation of two-polaron bound states and it has been shown that the helix supports two kinds of bound states referring to the trapping of two polarons onto the same amide-I mode and onto two nearest neighbor amide-I vibrations. These results were corroborated by a recent experiment devoted to the femtosecond infrared pump-probe spectroscopy of the N-H mode in a stable $\alpha$-helix which has revealed the two excited state absorption bands connected to the two kinds of bound states \cite{kn:hamm1,kn:hamm2}. Vibrational self-trapping in model helix was thus observed for the first time, validating in the same time the polaron approach.

However, most of the theoretical studies applied to the Davydov problem involve a 1D approximation of the 3D structure of the $\alpha$-helix. Although a real helix is formed by three spines of hydrogen-bonded peptide units connected through covalent bonds, such an assumption was motivated by the fact that when a vibron is symmetrically excited over the three spines, the vibron-phonon dynamics reduces to that of a single hydrogen bonded chain \cite{kn:scott1,kn:scott2,kn:scott3,kn:scott4}. However, it has been shown within the soliton approach, that when a vibron is excited on a single spine the 3D nature of the helix manifests itself by reducing the soliton velocity and by favoring the exchanges between spines \cite{kn:scott2,kn:scott3,kn:scott4}. Recently, a detailed analysis of both the soliton trapping and the soliton dynamics in a 3D $\alpha$-helix has been done by Hennig \cite{kn:hennig}. It has been shown that the soliton envelop exhibits a monotonic decay but shows a multihump structure. In addition, the lattice distortion along hydrogen bonds is reduced for the 3D model when compared with the 1D model, and, in the excited region, the radius of the helix decreases. From a dynamical point of view, Hennig has recovered the soliton velocity decrease and he has shown that the preferred transport pathway takes place along the hydrogen bonds.      

In this paper, the small polaron formalism is applied to a 3D model of an $\alpha$-helix with a special emphasis onto the modification of the dressing mechanism induced by the 3D nature of the phonons of the helix. We restrict our attention to the characterization of the single-polaron energy spectrum and the generalization to two-polaron eigenstates is addressed in the second paper of this series \cite{kn:helixII}. 

The paper is organized as follows. The model to describe the vibron-phonon dynamics in 3D $\alpha$-helices is introduced in Sec. II. In Sec. III, the nature of the phonons in the helix is first summarized. Then, we first realize a unitary transformation to remove the intramolecular anharmonicity and a Lang-Firsov \cite{kn:lang} transformation is applied to renormalize the vibron-phonon interaction. Finally, a mean field procedure is performed to obtain the effective dressed vibron Hamiltonian. In Sec. IV, a numerical analysis of the vibron-phonon dynamics is performed depending on the values of the relevant parameters of the problem. Finally, the results are discussed and interpreted in Sec. V. 

\section{Helix structure and Model Hamiltonian}
 
Let us consider a sequence of $N$ amino acid units (called residues) regularly distributed along a polypeptide chain. Let $n$ the index which labels the position of the $n$th residue along the chain. In three dimension, the structure of the polypeptide chain is 
stabilized by the hydrogen bonds between the carboxyl oxygen (CO) of residue $n$ and the amide hydrogen (NH) of residue $n+3$. The resulting conformation is a 3D $\alpha$-helix 
in which each residue is related to the next one by a translation of $h=1.5$ \AA\ and by a rotation of $\theta_{0}=100^{o}$ leading to 3.6 residues per turn of helix (see Fig. 1a). The pitch of the helix, i.e. the product of the translation (1.5 \AA) by the number of residues per turn ($3.6$), is equal to 5.4 \AA. The radius of the helix, which corresponds to the distribution of the center of mass of the residues, is fixed to $R_{0}=2.8$ \AA\ \cite{kn:davydov1}.

\begin{figure}
\includegraphics{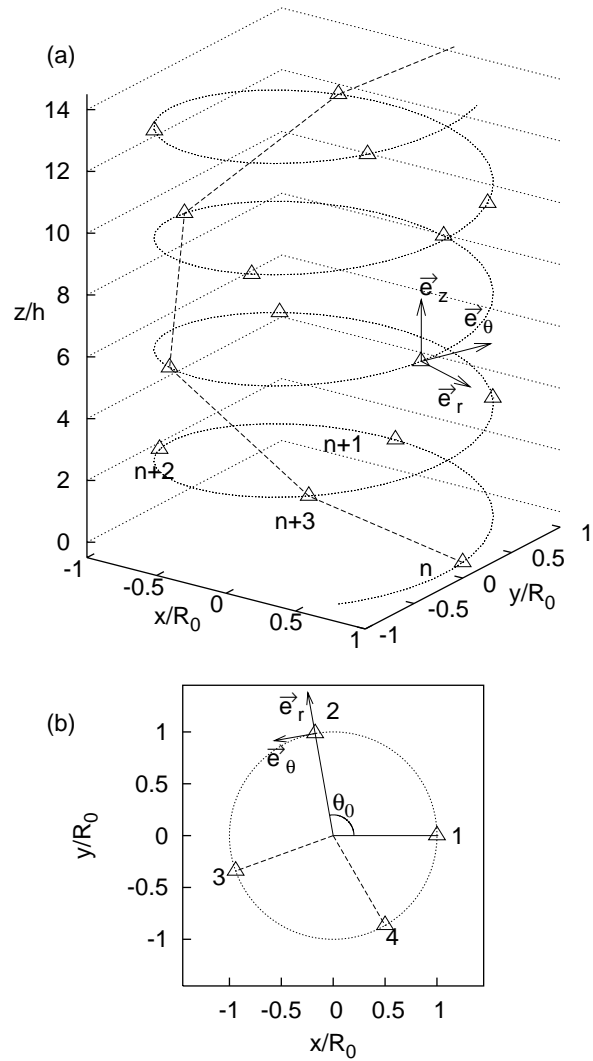}
\caption{(a) Representation of the $\alpha$-helix backbone defined by the angle $\theta_{0}=100^{o}$, the translation $h=1.5$ \AA\ and the radius $R_{0}=2.8$ \AA. Each site $n$ is occupied by a residue and the dashed line characterizes a particular spine of hydrogen-bonded peptide units. (b) Top view of the helix backbone where a particular local frame is displayed.}
\label{fig:helix}
\end{figure}

To model the vibron-phonon dynamics in a rather simple way, we follow the procedure introduced by Hennig \cite{kn:hennig} and treat the $N$ residues as point-like entities which the equilibrium positions are located on sites distributed along the helix. Therefore, as shown in Fig. 1a, the geometry of the helix is specified in the (xyz) cartesian frame where the $z$ direction is parallel to the axis of the helix. The equilibrium position of the $n$th residue is thus expressed as
\begin{equation}
\bm{R}(n) = R_0 \cos{n\theta_0} \bm{e_x} + R_0 \sin{n\theta_0} \bm{e_y} + nh \bm{e_z}
\label{eq:position}
\end{equation}  
where $\bm{e_\alpha}$ denotes the unit vector parallel to the direction $\alpha=x,y,z$.

The $n$th site contains a amide-I vibration which behaves as an internal high frequency oscillator described by the standard creation and annihilation vibron operators $b^{+}_{n}$ and $b_{n}$. By introducing the intramolecular anharmonicity of each amide-I mode, the vibron Hamiltonian is written as 
\begin{eqnarray} 
H_{v}&=& \sum_{n} \hbar\omega_{0}b^{\dag}_{n}b_{n}+\hbar\gamma_{3}(b^{\dag}_{n}+b_{n})^{3}+\hbar\gamma_{4}(b^{\dag}_{n}+b_{n})^{4}+...\nonumber \\
&-&\frac{1}{2} \sum_{n}\sum_{n'} \hbar J(n-n') (b^{\dag}_{n}+b_{n})(b^{\dag}_{n'}+b_{n'})
\label{eq:Hv}
\end{eqnarray}
where $\omega_{0}$ stands for the internal frequency of the $n$th amide-I mode and where $J(n-n')$ denotes the lateral hopping constant between the residues $n$ and $n'$. In Eq.(\ref{eq:Hv}), $\gamma_{3}$ and $\gamma_{4}$ represent the cubic and quartic anharmonic parameters of each amide-I mode. Note that for the single-vibron dynamics, the intramolecular anharmonicity does not play a crucial role. As it will be shown in the following of the text, it is responsible for a renormalization of the harmonic parameters (internal frequency, hopping constants, vibron-phonon coupling) but it does not fundamentally change the physical behavior of the single-vibron states. Nevertheless, it has been introduced here because it drastically affects the two-vibron dynamics, as shown in paper II \cite{kn:helixII}. 

The amide-I vibrations interact with the phonons of the helix associated to the collective dynamics of the external motions of the peptide groups. 
As mentioned by Hennig \cite{kn:hennig}, both the covalent and the hydrogen bonds are assumed to be represented by point-point interactions. The coupling between two residues is thus described by a pair potential $V[r(nn')]$ which depends on the instantaneous distance $ r(nn')=\mid \bm{r}(n)-\bm{r}(n')\mid $ between the two residues $n$ and $n'$. Within the harmonic approximation, each peptide group performs small displacements $\bm{u}(n)$ around its equilibrium position $\bm{R}(n)$ so that the instantaneous position $\bm{r}(n)$ of the $n$th residue can be expressed as $\bm{r}(n)=\bm{R}(n)+\bm{u}(n)$. Therefore, by expanding the intermolecular potential around the equilibrium helix conformation, the phonon Hamiltonian is written as
\begin{equation}
H_{p}=\sum_{n\alpha}\frac{p^{2}_{\alpha}(n)}{2M}+\frac{1}{2}\sum_{n\alpha,n'\beta} \Phi_{\alpha\beta}(nn')u_{\alpha}(n)u_{\beta}(n')
\label{eq:Hp}
\end{equation}
where $M$ denotes the average mass of each residue and where $\bm{p}(n)$ is the momentum associated to the displacement $\bm{u}(n)$. Both $\alpha$ and $\beta$, equal to $x,y$ or $z$, account for the representation of the vectors in the cartesian frame $(x,y,z)$.
In Eq.(\ref{eq:Hp}), $\bm{\Phi}(nn')$ is the force constant tensor defined as the second derivative of the potential at equilibrium and written as 
\begin{equation}
\Phi_{\alpha\beta}(nn')=\frac{\partial^{2}V[r(nn')]}{\partial u_{\alpha}(n)\partial u_{\beta}(n')}
\label{eq:PHI}
\end{equation}

According to the Davydov model, the vibron-phonon interaction originates in the modulation of the vibrational frequency of each amide-I vibration by the external motion of the residues. This coupling is expressed in terms of the instantaneous positions $\bm{r}(n)$ of the residues as 
\begin{equation}
\Delta H_{vp}=\sum_{n} \frac{1}{2} \delta h(\{ r(nn') \})(b^{\dag}_{n}+b_{n})^{2}
\end{equation}
Within the deformation potential approximation, the function $\delta h(\{ r(nn') \})$ can be expanded around the equilibrium position of the residues so that the resulting coupling Hamiltonian is finally expressed as 
\begin{equation}
\Delta H_{vp}=\sum_{n,n'}\frac{\chi_{\mid n-n' \mid}}{2}\frac{\bm{R}(nn') (\bm{u}_{n}-\bm{u}_{n'})}{R(nn')} (b^{\dag}_{n}+b_{n})^{2}
\label{eq:DHvp}
\end{equation}
where $\bm{R}(nn')=\bm{R}(n)-\bm{R}(n')$ and  where $\chi_{\mid n-n' \mid}$ denotes the strength of the interaction between the $n$th amide-I mode and the external displacement of the $n'$th residue. 

Finally, the vibron-phonon dynamics is described by the full Hamiltonian $H=H_v+H_p+\Delta H_{vp}$ which appears as a generalization of the well-known Davydov Hamiltonian to a 3D $\alpha$-helix.

\section{Theoretical Background}

In helices, the vibron-phonon interaction is assumed to be strong enough so that the vibron dynamics is essentially governed by the so-called dressing effect. The characterization of this effect requires the knowledge of the phonon eigenstates and of the Lang-Firsov transformation \cite{kn:lang} which yields the small polaron point of view. However, although these two features are well known in an infinite lattice with translational invariance, the helical nature of the lattice is responsible for a modification of both the phonon structure and the dressing mechanism. The present section is thus devoted to the presentation of these modifications. 

\subsection{Phonons in a 3D $\alpha$-helix}

In a recent paper devoted to solitons in an isolated helix chain, Christiansen et al. have introduced a general formalism to characterize the phonon dynamics in an $\alpha$-helix \cite{kn:chris2}. This formalism consists in expressing the residue dynamics in a simple way by taking advantage of the invariance of the phonon Hamiltonian according to a rotation by an angle of $\theta_0$ around the helix axis followed by a translation along this axis by a distance $h$ \cite{kn:fanconi}.  

To proceed, it is more convenient to express the position of the residues in a set of $N$ local frames $(\bm{e}_{r}(n),\bm{e}_{\theta}(n),\bm{e}_{z}(n))$ attached to each residue (see Fig. 1b). The unit vectors of the $n$th local frame are expressed in terms of the unit vectors of the cartesian frame $(\bm{e}_{x},\bm{e}_{y},\bm{e}_{z})$ as 
\begin{equation}
\bm{e}_{\mu}(n)=T_{\mu\alpha}(n)\bm{e}_{\alpha}
\end{equation}
where $\mu=r,\theta,z$ and $\alpha=x,y,z$ and where $\bm{T}(n)$ is the local $(3\times3)$ rotational matrix defined as
\begin{equation}
\bm T (n)=
\begin{pmatrix}
\cos{n\theta_0} & \sin{n\theta_0} & 0 \\
-\sin{n\theta_0} & \cos{n\theta_0} & 0 \\
0 & 0 & 1 
\end{pmatrix}
\label{eq:T}
\end{equation}

In that context, let $\bm{v}(n)$ denotes the displacement of the $n$th residue expressed in the $n$th local frame. This vector is defined in terms of its cartesian counterpart $\bm{u}(n)$ as $\bm{v}(n)=\bm{T}(n)\bm{u}(n)$. Therefore, by applying this latter transformation, the phonon Hamiltonian Eq.(\ref{eq:Hp}) is rewritten as
\begin{equation}
H_{p}=\sum_{n\mu}\frac{p^{2}_{\mu}(n)}{2M}+\frac{1}{2}\sum_{n\mu,n'\nu} \Psi_{\mu\nu}(nn')v_{\mu}(n)v_{\nu}(n')
\label{eq:HP}
\end{equation}
where $p_{\mu}(n)$ denotes the momentum connected to $v_{\mu}(n)$ and where $\bm{\Psi}(nn')=\bm{T}(n)\bm{\Phi}(nn')\bm{T}^{-1}(n')$.

At this step, it is straightforward to show that the transformed force constant tensor $\bm{\Psi}(nn')$ depends only on the index difference $n-n'$ (see Appendix A). As a consequence, the Bloch theorem can be applied so that the  displacement of the $n$th residue in the $n$th local frame can be expanded as a superimposition of plane waves as
\begin{equation}
v_{\mu}(n)=\frac{1}{\sqrt{N}}\sum_{k}v_{\mu}(k)e^{ikn}
\end{equation}
where $k$ denotes a reduced wave vector which belongs to the first Brillouin zone of the helix, i.e. $-\pi<k<\pi$ . The phonon Hamiltonian is thus defined in terms of the $(3 \times 3)$ dynamical matrix $D_{\mu\nu}(k)=\sum_{n}\Psi_{\mu\nu}(0n)\exp(ikn)/M$  as
\begin{equation}
H_{p}=\sum_{k\mu}\frac{p^{2}_{\mu}(k)}{2M}+\frac{M}{2}\sum_{k\mu,\nu} D_{\mu\nu}(k)v_{\mu}(k)v_{\nu}(k)
\end{equation}
where $\bm{p}(k)$ is the momentum associated to $\bm{v}(k)$. Note that the expression of the Dynamical matrix is detailed in Appendix A.

Finally, the normal mode decomposition is achieved by performing the diagonalization of the dynamical matrix $\bm{D}(k)$ for each $k$ value. Such a procedure allows us to define three eigenvalues $\Omega^{2}_{k\sigma}$ and three eigenvectors $\bm{\epsilon}_{k\sigma}$ labeled by the index $\sigma=1,2,3$. Therefore, the two indexes $k$ and $\sigma$ specify a particular phonon mode with energy $\hbar\Omega_{k\sigma}$, quasi-momentum $\hbar k$ and polarization $\bm{\epsilon}_{k\sigma}$. The quantum dynamics of each mode is described by the well-known creation $a^{\dag}_{k\sigma}$ and annihilation $a_{k\sigma}$ operators so that the phonon Hamiltonian can be rewritten in the standard form as 
\begin{equation}
H_{p}=\sum_{k\sigma} \hbar \Omega_{k\sigma} (a^{\dag}_{k\sigma}a_{k\sigma}+\frac{1}{2})
\end{equation}
In the same way, the displacement of the $n$th residue is finally expressed as 
\begin{equation}
v_{\mu}(n)=\sum_{k\sigma} \sqrt{\frac{\hbar}{2MN\Omega_{k\sigma}}}(a_{k\sigma}+a^{\dag}_{-k\sigma})\epsilon_{\mu k\sigma}e^{ikn}
\label{eq:vmu}
\end{equation}

At this step, the vibron-phonon coupling Hamiltonian Eq.(\ref{eq:DHvp}) can be rewritten in terms of the phonon normal mode coordinates as
\begin{equation}
\Delta H_{vp}=\sum_{k \sigma} \hbar(\frac{\Delta_{k\sigma}}{2}e^{-ikn} a^{\dag}_{k\sigma}+\frac{\Delta^{*}_{k\sigma}}{2}e^{ikn} a_{k\sigma})(b^{\dag}_{n}+b_{n})^{2}
\label{eq:DHvp2}
\end{equation}
where $\Delta_{k\sigma}$ accounts for the modulation of the frequency of the $n$th amide-I vibration due to its coupling with the phonon mode specified by the wave vector $k$ and the index $\sigma$. It is defined as
\begin{equation}
\Delta_{k\sigma}= \sum_{m}  \frac{\chi_{\mid m \mid}\bm{C}_{k}(m)\bm{\epsilon}^{*}_{k\sigma}}
{\sqrt{2MN\hbar \Omega_{k\sigma}}}
\label{eq:DELTA}
\end{equation}
where $\bm{C}_{k}(m)$ denotes a vector expressed as
\begin{equation}
\bm{C}_{k}(n'-n)=[\bm{T}(n)-\bm{T}(n')e^{ik(n-n')}]\frac{\bm{R}(nn')}{R(nn')}
\end{equation}

\subsection{Anharmonic vibron, small polaron and effective Hamiltonian}

To remove the cubic and quartic intramolecular anharmonicity, the standard procedure described by Kimball et al. \cite{kn:kimball} is used. First, by disregarding the lateral coupling between nearest neighbor amide-I vibrations as well as the vibron-phonon interaction, the procedure consists in performing a unitary transformation $V$ to diagonalize each anharmonic amide-I mode. Then, an approximate Hamiltonian is obtained by applying the transformation on the full Hamiltonian $H$ and be keeping the vibron-conserving terms, only. As a result, by using the procedure detailed in Ref. \cite{kn:pouthier1,kn:pouthier2}, the transformed Hamiltonian $\tilde{H}=VHV^{+}$ is expressed as (within the convention $\hbar$=1)

\begin{eqnarray}
\tilde{H}&=&\sum_{n} (\omega_{0}-2A-B)b^{\dag}_{n}b_{n}-\sum_{nn'}J_{1}(n-n')b^{\dag}_{n}b_{n'}+H_{p} \nonumber \\
&+&\sum_{k \sigma} (1+2\eta)(\Delta_{k\sigma}e^{-ikn} a^{\dag}_{k\sigma}+\Delta^{*}_{k\sigma}e^{ikn} a_{k\sigma})b^{\dag}_{n}b_{n}
\label{eq:HTILDE}
\end{eqnarray}
where $A=30\gamma_{3}^{2}/\omega_{0}-6 \gamma_{4}$ denotes the positive anharmonic parameter and where the other parameters in Eq.(\ref{eq:HTILDE}) are defined as 
\begin{eqnarray}
B&=&\sum_{n}72J(n)(\gamma_{3}/\omega_{0})^{2} \nonumber \\
J_{1}(n)&=&J(n)(1+44(\gamma_{3}/\omega_{0})^{2}-12\gamma_{4}/\omega_{0}) \nonumber \\
\eta&=&120(\gamma_{3}/\omega_{0})^{2}-12\gamma_{4}/\omega_{0}
\end{eqnarray}

Note that since we restrict our attention to the single-vibron dynamics, additional terms which involve the product of four vibron operators have been disregarded in the present version of the Hamiltonian Eq.(\ref{eq:HTILDE}). These terms play a crucial role for the two-vibron dynamics \cite{kn:pouthier1,kn:pouthier2} and they are addressed in the second paper of this series \cite{kn:helixII}. 

To partially remove the vibron-phonon coupling Hamiltonian, a Lang-Firsov transformation is applied \cite{kn:lang}. Indeed, since the vibron-phonon dynamics is dominated by the so-called dressing effect, we consider a "full dressing" and introduce the following unitary transformation
\begin{equation}
U=\exp(\sum_{nk\sigma}(1+2\eta)[\frac{\Delta_{k\sigma}e^{-ikn}}{\Omega_{k\sigma}}a^{\dag}_{k\sigma}-\frac{\Delta^{*}_{k\sigma}e^{ikn}}{\Omega_{k\sigma}}a_{k\sigma}]b_{n}^{\dag}b_{n})
\label{eq:U}
\end{equation}
By using Eq.(\ref{eq:U}), the transformed Hamiltonian $\hat{H}=U \tilde{H} U^\dag$
is written as 
\begin{eqnarray}
\hat{H}&=& H_{p}+\sum_{n}(\omega_{0}-2A-B-(1+4\eta)E_{B})b^{\dag}_{n}b_{n} \nonumber \\ 
&-&\sum_{nn'}J_{1}(n-n')\Theta_{n}^\dag\Theta_{n'}b^\dag_{n}b_{n'}
\label{eq:HHAT}
\end{eqnarray}
where $E_{B}$ denotes the small polaron binding energy defined as
\begin{equation}
E_{B}=\sum_{k\sigma} \frac{\mid \Delta_{k\sigma} \mid^{2} }{\Omega_{k\sigma}}
\end{equation}
whereas $\Theta_{n}$ stands for the dressing operator expressed as
\begin{equation}
\Theta_{n}=\exp(-\sum_{k\sigma}(1+2\eta)[\frac{\Delta_{k\sigma}e^{-ikn}}{\Omega_{k\sigma}}a^{\dag}_{k\sigma}-\frac{\Delta^{*}_{k\sigma}e^{ikn}}{\Omega_{k\sigma}}a_{k\sigma}])
\label{eq:theta}
\end{equation}

In this dressed vibron point of view (Eq.(\ref{eq:HHAT})), the vibron-phonon coupling remains through the modulation of the lateral terms by the dressing operators. Although these operators depend on the phonon coordinates in a highly nonlinear way, the vibron-phonon interaction has been strongly reduced within this transformation. As a result, we can take advantage of such a reduction to perform a mean field procedure \cite{kn:ivic1,kn:ivic2,kn:yarkony} and to express the full Hamiltonian $\hat{H}$ as the sum of three separated contributions as 
\begin{equation}
\hat{H}=\hat{H}_{eff}+H_{p}+\Delta H 
\end{equation}
where $\hat{H}_{eff}=\langle(\hat{H}-H_{p})\rangle$ denotes the effective Hamiltonian of the dressed vibrons and where $\Delta H =\hat{H}-H_{p}-\langle (\hat{H}-H_{p})\rangle$ stands for the remaining part of the vibron-phonon interaction. The symbol $\langle\hdots\rangle$ represents a thermal average over the phonon degrees of freedom which are assumed to be in thermal equilibrium at temperature $T$.

As a result, the effective dressed vibron Hamiltonian is written as
\begin{equation}
\hat{H}_{eff}=\sum_{n}\hat{\omega}_{0}b^{\dag}_{n}b_{n}-\sum_{nn'} J_{eff}(n-n')b^\dag_{n}b_{n'} 
\label{eq:HHATEFF}
\end{equation}
where $\hat{\omega}_{0}=\omega_{0}-2A-B-(1+4\eta)E_{B}$ and where the effective vibron hopping constant $J_{eff}(n-n')=J_{1}(n-n')\exp(-(1+4\eta)S(n-n',T))$ is expressed in terms of the coupling constant $S(n,T)$ defined as 
($k_{B}$ denotes the Boltzmann constant)
\begin{equation}
S(n,T)=\sum_{k\sigma} \mid \frac{\Delta_{k\sigma}}{\Omega_{k\sigma}} \mid^2 \coth(\frac{\hbar \Omega_{k\sigma}}{2k_BT})(1-\cos(kn))
\label{eq:S}
\end{equation}

The Hamiltonian $\hat{H}_{eff}$ (Eq.(\ref{eq:HHATEFF})) describes the dynamics of anharmonic vibrons dressed by a virtual cloud of phonons, i.e. small polarons. A polaron state corresponds to a delocalized plane wave with wave vector $q$ and frequency $\omega_{q}=\hat{\omega}_{0}-2\sum_{n\geq1}J_{eff}(n)\cos(qn)$. Eq.(\ref{eq:HHATEFF}) accounts for a renormalization of the main part of the vibron-phonon coupling within the non-adiabatic limit. The remaining vibron-phonon coupling is thus assumed to be small in order to be treated using perturbative theory. Nevertheless, such a contribution is disregarded in the present work.

\section{Numerical results}

In this section, the previous formalism is applied to characterize the single-vibron energy spectrum of a 3D $\alpha$-helix. However, the present theory involves a set of parameters which has first to be discussed. 

From the literature, the parameters enter the vibron dynamics in 3D $\alpha$-helices are relatively well described . The quantum energy for an amide-I vibration is about 1660 cm$^{-1}$ so that the harmonic frequency is fixed to $\omega_{0}=1695$ cm$^{-1}$. A well admitted value for the hopping constant along the hydrogen bonds is $J(3)=7.8$ cm$^{-1}$ (see for instance Refs. \cite{kn:scott1,kn:ivic2}). In the same way, the vibron hopping constants between different spines of hydrogen-bonded peptide units have been calculated in Ref. \cite{kn:scottA} so that $J(1)=-12.4$ cm$^{-1}$ and $J(2)=3.9$ cm$^{-1}$. Despite their small values, the other elements of the hopping matrix have been taken into account in the present work and are listed in the Table I of Ref. \cite{kn:scottA}. Finally, the anharmonic constant of the amide-I vibration is about $A=8.0$ cm$^{-1}$ \cite{kn:pouthier1,kn:hamm3,kn:hamm4} and both the cubic and quartic anharmonic parameters can be expressed approximately by using the relation $15\gamma_{3}^{2}/\omega_{0} \approx 6 \gamma_{4}\approx A$ \cite{kn:pouthier1}. Note that these parameter values
are in a rather good agreement with recent ab initio calculations \cite{kn:muk}.

The phonon dynamics is essentially governed by the force constant tensor defined in Eq.(\ref{eq:PHI}). In that context, we assume that the helical conformation corresponds to the equilibrium of each pair potential which connects two residues. Therefore, the force constant tensor is expressed as
\begin{equation}
\Phi_{\alpha \beta}(nn')=-K_{\mid n-n'\mid}\frac{R_{\alpha}(nn')R_{\beta}(nn')}{R^{2}(nn')}
\end{equation}
where $K_{\mid n-n'\mid}$ denotes the second derivative of the pair potential connecting the two residues located onto the sites $n$ and $n'$. 
According to the 3D model of Hennig \cite{kn:hennig}, the $\alpha$-helix exhibits two modes corresponding to a longitudinal displacement along the helix axis and to a radial displacement perpendicular to this axis. To recover these results, the force constant $K_{2}$ is set to zero and we assume $K_{\mid n \mid}=0$ for $n\geq4$. 
The force constant $K_{3}$ of the hydrogen bonds ranges between 13 and 20 Nm$^{-1}$ \cite{kn:scott1,kn:ivic2,kn:pouthier1,kn:pouthier2,kn:hennig} whereas the force constant $K_{1}$ of the covalent bonds varies between 45 and 75 Nm$^{-1}$ \cite{kn:hennig}. With our notations, $K_{1}\approx W/2$, where $W$ is the force constant for the radial motion introduced by Hennig. Finally, the mass $M$ which enters the phonon dynamics has been fixed to 2.0 $10^{-25}$ kg. 

The vibron-phonon coupling is basically defined from the knowledge of the parameters $\chi_{\mid n-n'\mid}$ (Eq.(\ref{eq:DHvp})) which accounts for the modulation of the $n$th amide-I frequency due to the external motion of the $n'$th residue. According to the 1D Davydov model, the parameter $\chi_{3}$ ranges between 35 and 62 pN. 
However, the influence of the motion of the residues $n\pm1$ and $n\pm2$ onto the $n$th amide-I vibration is still unknown. Therefore, we choose $\chi_{2}=0$ whereas we treat $\chi_{1}$ as a parameter smaller than $\chi_{3}$ \cite{kn:hennig}. Note that $\chi_{n}=0$ for $n \geq 4$.

\begin{figure}
\includegraphics{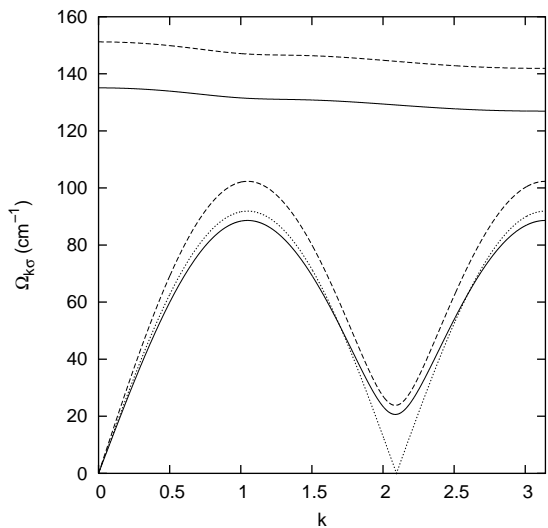}
\caption{Phonon dispersion curves for $K_{1}=60$ Nm$^{-1}$, $K_{3}=15$ Nm$^{-1}$ (full line) and for $K_{1}=75$ Nm$^{-1}$, $K_{3}=20$ Nm$^{-1}$ (dashed line) (Note that $K_{2}=0$). The dotted line represents the phonon dispersion curve of the 1D Davydov model with $K_{3}=15$ Nm$^{-1}$.}
\end{figure}

In Fig. 2, the phonon energy spectrum is shown for $K_{1}=60$ Nm$^{-1}$, $K_{3}=15$ Nm$^{-1}$ (full line) and for $K_{1}=75$ Nm$^{-1}$, $K_{3}=20$ Nm$^{-1}$ (dashed line).
The spectrum exhibits two non vanishing dispersion curves whereas the third one is equal to zero in the entire  Brillouin zone. The low frequency curve, which tends to zero when $k\rightarrow 0$, corresponds to acoustical phonons, whereas the high frequency curve represents optical phonons. 
In the long-wavelength limit, the acoustic branch increases linearly with the wave vector $k$ and describes a sound wave. However, the curve reaches a maximum value when $k=\pi/3$ then it decreases to a minimum value when $k=2\pi/3$. Finally, the frequency increases again when $k$ reaches the edge of the Brillouin zone. Note that the acoustic branch is rather insensitive to the force constant $K_{1}$ whereas it strongly depends on the force constant $K_{3}$. The maximum acoustic frequencies are equal to 89 and 102 cm$^{-1}$ when $K_{3}=15$ and $20$ Nm$^{-1}$, respectively, whereas the minimum frequency in $k=2\pi/3$ is about 20 cm$^{-1}$. The optical branch appears almost constant over the entire Brillouin zone and exhibits a rather weak dispersion of about 8.5 cm$^{-1}$. It strongly depends on the force constant $K_{1}$ and appears almost insensitive to $K_{3}$. At the center of the Brillouin zone, the optical frequency is equal to 135 cm$^{-1}$ when $K_{1}=60$ Nm$^{-1}$ and reaches 151 cm$^{-1}$ when $K_{1}=75$ Nm$^{-1}$. Note that the dotted line represents the phonon dispersion curve in the corresponding 1D Davydov model for $K_{3}=15$ Nm$^{-1}$. 

In the long-wavelength limit, the analysis of the eigenvectors of the dynamical matrix reveals that the acoustic phonons are essentially polarized along the axis of the helix. More precisely, they correspond to longitudinal vibrations along the direction specified by the hydrogen bonds. By contrast, the optical branch refers to a radial motion of the residues which describes a breathing mode of the radius of the helix. These two features corroborate the previous observed dependence of the two branches with respect to the force constants $K_{1}$ and $K_{3}$. Note that a singular behavior takes place when $k=2\pi/3$ since the acoustic branch becomes polarized along the radial direction, only. In the same way, when $k$ reaches the edge of the Brillouin zone, the optical branch refers to an hybridization between a radial and a torsional motion. 

\begin{figure}
\includegraphics{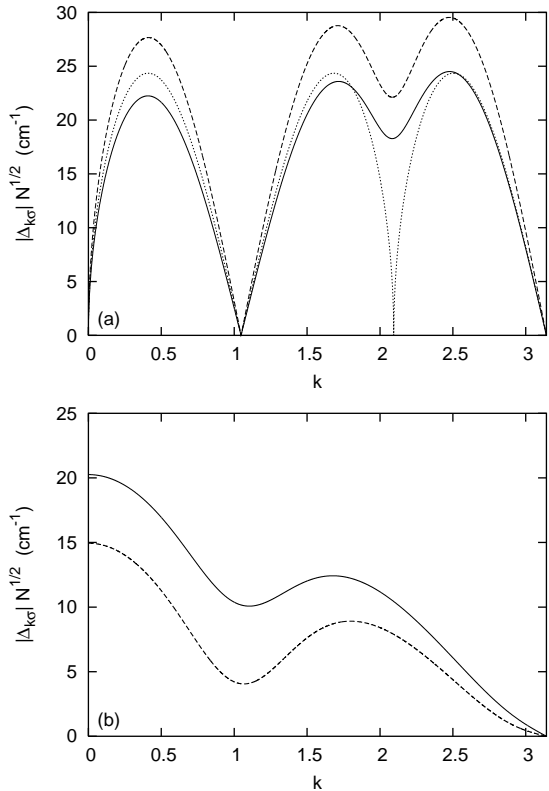}
\caption{Normalized function $\sqrt{N}\mid \Delta_{k\sigma}(n) \mid$ vs $k$ for $K_{1}=60$ Nm$^{-1}$, $K_{3}=15$ Nm$^{-1}$ and for $\chi_{1}=25$ pN, $\chi_{3}=50$ pN (full line) and $\chi_{1}=10$ pN, $\chi_{3}=60$ pN (dashed line). (a) coupling with acoustical phonons, (b) coupling with optical phonons. The dotted line in Fig. 3a refers to the 1D Davydov model.}
\end{figure}

To characterize the vibron-phonon interaction, the $k$ dependence of the normalized function $\sqrt{N}\mid \Delta_{k\sigma} \mid$ defined in Eq.(\ref{eq:DELTA}) is displayed in Figs. 3.
The phonon force constants are fixed to $K_{1}=60$ Nm$^{-1}$ and $K_{3}=15$ Nm$^{-1}$ and two situations are considered for the strength of vibron-phonon coupling, i.e $\chi_{1}=25$ pN, $\chi_{3}=50$ pN (full line) and $\chi_{1}=10$ pN, $\chi_{3}=60$ pN (dashed line). 
As shown in Fig. 3a, the coupling with acoustical phonons strongly depends on the phonon wave vector. It vanishes for three $k$ values defined as $k^{(1)}=0$, $k^{(2)}=\pi/3$ and $k^{(3)}=\pi$. The coupling exhibits three maxima located at $k'^{(1)}=0.41$, $k'^{(2)}=1.71$ and $k'^{(3)}=2.48$. When $\chi_{1}=25$ pN and $\chi_{3}=50$ pN, the maxima lye around 23 cm$^{-1}$ whereas they reach 28 cm$^{-1}$ when $\chi_{1}=10$ pN and $\chi_{3}=60$ pN. In fact, the coupling with acoustical phonon depends significantly on $\chi_{3}$ and appears almost insensitive to $\chi_{1}$. Note that the coupling shows a local minimum around $k=2\pi/3$ which correspond to a zero coupling in the 1D Davydov model (dotted line). 

As shown in Fig. 3b, the coupling with optical phonons basically decreases when the phonon wave vector increases. It is maximum at the center of the Brillouin zone and vanishes when $k=\pi$. When $\chi_{1}=25$ pN and $\chi_{3}=50$ pN, the maximum is equal to 20 cm$^{-1}$ whereas it reduces to 15 cm$^{-1}$ when $\chi_{1}=10$ pN and $\chi_{3}=60$ pN. In a marked contrast with the previous situation, the coupling with optical phonon depends essentially on $\chi_{1}$. Note that this coupling shows a local minimum around $k=\pi/3$.

\begin{figure}
\includegraphics{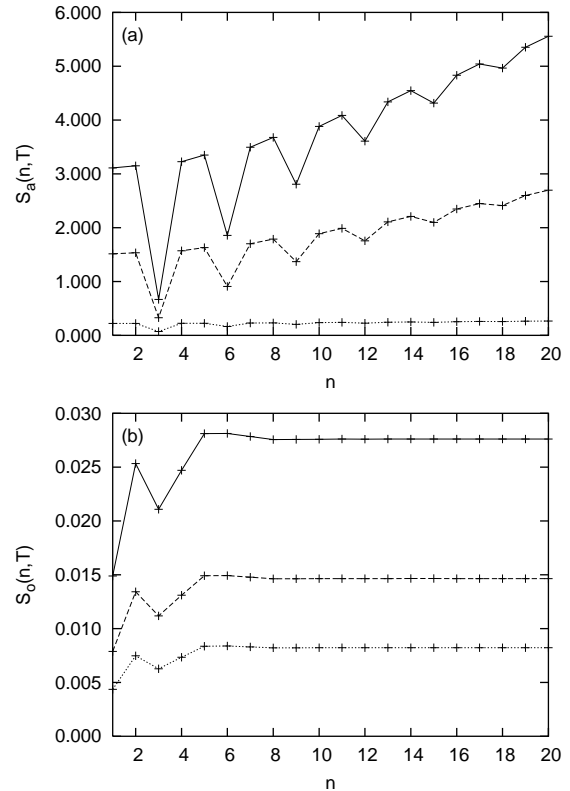}
\caption{Coupling constant $S(n,T)$ vs the bond index $n$ for $T=310$ K, (full line), $T=150$ K (dashed line) and $T=5$ K (dotted line). The phonon force constants are fixed to $K_{1}=60$ Nm$^{-1}$ and $K_{3}=15$ Nm$^{-1}$ and the vibron-phonon coupling constants are equal to $\chi_{1}=25$ pN and $\chi_{3}=50$ pN. (a) Contribution of the acoustical phonons, (b) contribution of the optical phonons.}
\end{figure}

In Figs. 4, the coupling constant $S(n,T)$ defined in Eq.(\ref{eq:S}) is shown for $T=310$ K, (full line), $T=150$ K (dashed line) and $T=5$ K (dotted line) and for $K_{1}=60$ Nm$^{-1}$, $K_{3}=15$ Nm$^{-1}$, $\chi_{1}=25$ pN and $\chi_{3}=50$ pN. The sum over the index $\sigma$ in Eq.(\ref{eq:S}) allows us to express the coupling constant as $S(n,T)=S_{a}(n,T)+S_{o}(n,T)$ where $S_{a}(n,T)$ and $S_{o}(n,T)$ denote the contributions of the coupling with acoustical and optical phonons, respectively. 

As illustrated in Fig. 4a, at low temperature, the coupling constant $S_{a}(n,T)$ appears almost site independent and slightly fluctuates around 0.25. When the temperature increases, the coupling constant increases and exhibits two distinct behaviors depending on the nature of the bond $n$ which experiences the dressing effect. Indeed,   
when $n=3,6,9,...$, $S_{a}(n,T)$ accounts for the dressing mechanism responsible for a decreases of the vibron hopping constant between amide-I vibrations along the same spine of hydrogen-bonded peptide units. The coupling $S_{a}(n,T)$ increases almost linearly with $n$ and leads to a screening of the corresponding vibron hopping constants. When $n=1,2,4,5,...$, $S_{a}(n,T)$ characterizes the dressing of the hopping constant between amide-I vibrations located onto  different spines of hydrogen-bonded peptide units. As previously, it increases as $n$ increases. The figure clearly shows that vibron hops between different spines experience a stronger dressing mechanism than the hops which take place along the same spine. The difference between the two kinds of dressing increases as the temperature increases. However, as $n$ increases, the two kinds of dressing converge to the same $n$ dependence. In other words, long range vibron hops experience the same dressing effect whatever the location of the amide-I vibrations involved in the process. 

As shown in Fig. 4b, the contribution of the optical phonons $S_{o}(n,T)$ is about two orders of magnitude smaller than the contribution of acoustical phonons. It exhibits a fully different behavior with respect to the bond index $n$ since it drastically varies with $n$ until $n\leq4$ and becomes almost $n$ independent when $n\geq5$. Note that as previously, the coupling constant $S_{o}(n,T)$ increases with the temperature. 

\begin{figure}
\includegraphics{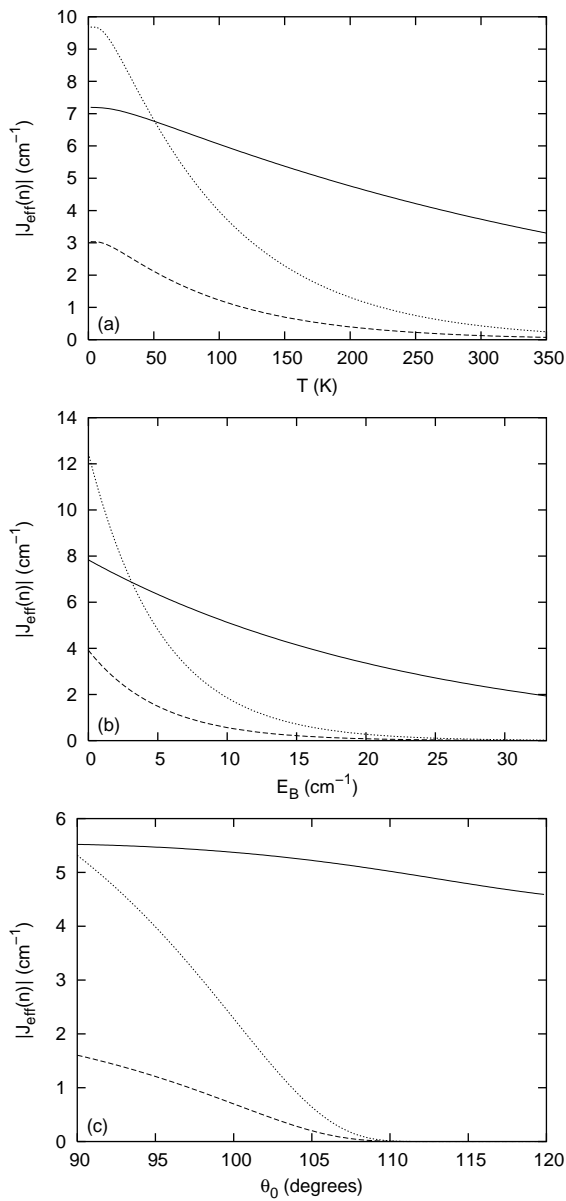}
\caption{Behavior of the effective vibron hopping constants $\mid J_{eff}(n) \mid$ for $n=1$ (dotted line), $n=2$ (dashed line) and $n=3$ (full line). The parameters used for the calculations are $K_{1}=60$ Nm$^{-1}$ and $K_{3}=15$ Nm$^{-1}$. (a) Temperature dependence for $\chi_{1}=25$ pN and $\chi_{3}=50$ pN. (b) $E_{B}$ dependence for $T=150$ K. In that case, $E_{B}$ varies according to the restriction $\chi_{1}=\chi_{3}/2$. (c) $\theta_{0}$ dependence for $T=150$ K.}
\end{figure}

The temperature dependence of the effective vibron hopping constants $\mid J_{eff}(n) \mid$ for $n=1$ (dotted line), $n=2$ (dashed line) and $n=3$ (full line) is drawn in Fig. 5a. The parameters used for the calculations are $K_{1}=60$ Nm$^{-1}$, $K_{3}=15$ Nm$^{-1}$, $\chi_{1}=25$ pN and $\chi_{3}=50$ pN. Due to the dressing effect, the three hopping constants decrease as the temperature increases. In addition, whatever the temperature, the hopping constant $ \mid J_{eff}(2) \mid$ is always smaller than the two other hopping constants. However, the ratio between $\mid J_{eff}(1) \mid$ and $\mid J_{eff}(3)\mid$ strongly depends on the temperature and a transition occurs when $T=T_{c}=51$ K. When $T<T_{c}$, the constant $\mid J_{eff}(1) \mid$ is the dominant contribution. For instance, at $T=0$ K, $J_{eff}(1)=-9.7$ cm$^{-1}$, $J_{eff}(2)=3.0$ cm$^{-1}$ and $J_{eff}(3)=7.2$ cm$^{-1}$. By contrast, when $T>T_{c}$, $\mid J_{eff}(3) \mid$ becomes the main hopping constant, and, for instance at $T=300$ K, $J_{eff}(1)=-0.43$ cm$^{-1}$, $J_{eff}(2)=0.13$ cm$^{-1}$ and $J_{eff}(3)=3.7$ cm$^{-1}$.
A similar behavior is illustrated in Fig. 5b for the dependence of the hopping constants with respect to the small polaron binding energy $E_{B}$. The temperature is fixed to $T=150$ K and $E_{B}$ varies according to the restriction $\chi_{1}=\chi_{3}/2$. Due to the dressing effect, the effective hopping constants decrease as $E_{B}$ increases. As previously, $\mid J_{eff}(2)\mid$ is smaller than the two other constants and a transition occurs when $E_{B}=3.1$ cm$^{-1}$ which discriminates between $\mid J_{eff}(1)\mid$ and $\mid J_{eff}(3)\mid$. Therefore, $\mid J_{eff}(1) \mid$ is greater than $\mid J_{eff}(3) \mid$ when $E_{B}<3.1$ cm$^{-1}$ whereas the opposite feature occurs when $E_{B}>3.1$ cm$^{-1}$. Note that when $E_{B}>25$ cm$^{-1}$, both $\mid J_{eff}(1)\mid$ and $\mid J_{eff}(2)\mid$ almost vanish so that only the hopping constant $\mid J_{eff}(3)\mid$ remains.
Finally, the influence of the angle $\theta_{0}$ on the vibron hopping constants is illustrated in Fig. 5c for $T=150$ K. The figure clearly shows that a change in $\theta_{0}$ is accompanied by drastic modifications in the hopping constants  $\mid J_{eff}(1) \mid$ and $\mid J_{eff}(2) \mid$ whereas $ \mid J_{eff}(3) \mid$ is just slightly modified. When $\theta_{0}$ is greater than 110$^{o}$, both $J_{eff}(1) $ and $ J_{eff}(2)$ vanish so that only vibron hops along a given spine of hydrogen-bond peptide units remains. 

\begin{figure}
\includegraphics{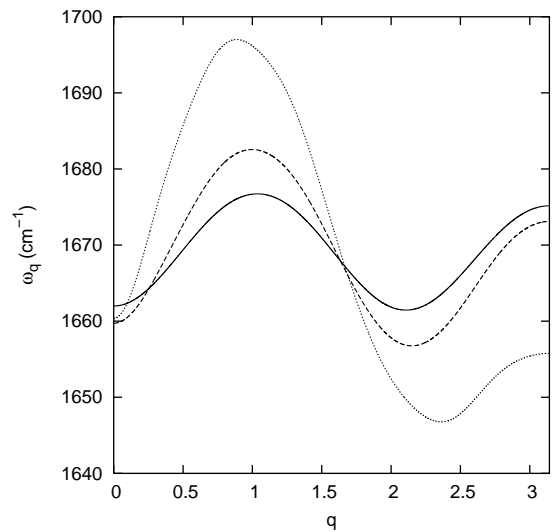}
\caption{Small polaron dispersion curve for $T=310 K$ (full line), $T=150 K$ (dashed line) and $T=5 K$ (dotted line) and for $K_{1}=60$ Nm$^{-1}$, $K_{3}=15$ Nm$^{-1}$, $\chi_{1}=25$ pN and $\chi_{3}=50$ pN. }
\end{figure}

The small polaron dispersion curve is displayed in Fig. 6 for $T=310 K$ (full line), $T=150 K$ (dashed line) and $T=5 K$ (dotted line) and for $K_{1}=60$ Nm$^{-1}$, $K_{3}=15$ Nm$^{-1}$, $\chi_{1}=25$ pN and $\chi_{3}=50$ pN. 
At the center of the Brillouin zone, the polaron frequency is almost temperature independent and is located around 1660 cm$^{-1}$. The dispersion curves exhibit a maximum around $q=1.0$ equal to 1697 cm$^{-1}$, 1683 cm$^{-1}$ and  1677 cm$^{-1}$ for $T=5$ K, $T=150$ K and $T=310$ K, respectively. In the same way, they show a minimum around $q=2.15$ equal to 1647 cm$^{-1}$, 1657 cm$^{-1}$ and  1661 cm$^{-1}$ for $T=5$ K, $T=150$ K and $T=310$ K, respectively. The polaron bandwidth, defined as the difference between the maximum frequency and the minimum frequency, decreases as the temperature increases. It is equal 16 cm$^{-1}$ for $T=310$ and reaches 50 cm$^{-1}$ for $T=5$ K. Note that the polaron frequency exhibits a strong temperature dependence at the edge of the Brillouin zone.  

\begin{figure}
\includegraphics{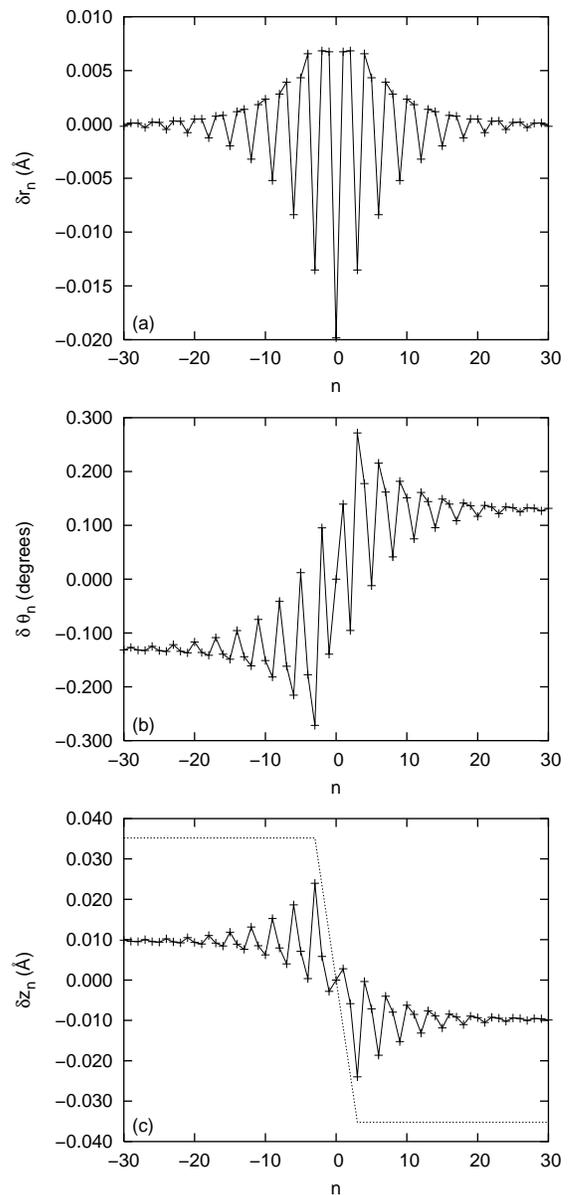}
\caption{Deformation of the helix backbone due to the creation of a vibron onto the site $n_{0}=0$ for $K_{1}=60$ Nm$^{-1}$, $K_{3}=15$ Nm$^{-1}$, $\chi_{1}=25$ pN and $\chi_{3}=50$ pN. (a) radial deformation, (b) angular torsion and (c) longitudinal distortion. The Dotted line in Fig. 7c represents the distortion in the 1D Davydov model.}
\end{figure}

As defined in Appendix B (see also Refs. \cite{kn:def1,kn:def2}), the distortion of the helix backbone due to the creation of a vibron onto the site $n_{0}=0$ is displayed in Figs. 7 for  $K_{1}=60$ Nm$^{-1}$, $K_{3}=15$ Nm$^{-1}$, $\chi_{1}=25$ pN and $\chi_{3}=50$ pN. Fig. 7a clearly shows that the vibron is responsible for a decrease of the radius of the helix onto the sites $n_{0}$, $n_{0}\pm3$, $n_{0}\pm6$ ... which are linked to each other via hydrogen bonds. By contrast, the radius of the helix on the sites $n_{0}\pm1$, $n_{0}\pm2$, $n_{0}\pm4$, $n_{0}\pm5$ ... , slightly increases but with smaller amplitudes. Note that the deformation of the radius is exponentially localized around the excited site $n_{0}$.   
The angular distortion of the helix which corresponds to a torsional motion of the residues is shown in Fig. 7b. Although the site $n_{0}$ does not exhibit any torsion, oscillations in the deformation of the helix backbone take place due to the different kinds of bonds between the site $n_{0}$ and its neighboring sites.  
Nevertheless, after the distortion has been propagated over few residues, the figure clearly shows that the residues located on the right side of the excited site $(n>n_{0})$ are characterized by an increase of the $\theta$ angle around its equilibrium value. By contrast, the opposite feature appears in the left side of the excited site $(n<n_{0})$ where the $\theta$ angle of each residue is basically reduced from its equilibrium value. As a result, a kink-like soliton takes place which discriminates between a negative torsion of about -0.12 degree and a positive torsion of about +0.12 degree.
Similar features are observed for the deformation along the axis of the helix as shown in Fig. 7c. Indeed, although the central site does not exhibit any distortion, a kink-like soliton takes place. In the left side of the excited side, the new equilibrium positions of the residues along the helix axis correspond to an increase of the $z$ coordinates whereas the opposite feature occurs in the right side of the excited site. The curve exhibits oscillations which clearly show that the distortion essentially affects the residues $n_{0}\pm3$, $n_{0}\pm6$, ... linked via hydrogen bonds. Note that the amplitude of the distortion is about 0.02 \AA\ and appears almost 3.5 times smaller than the deformation which occurs in the corresponding 1D Davydov model (dotted line). 

\section{discussion}

In the previous section, the numerical calculations have clearly shown that the 3D nature of the $\alpha$-helix strongly modifies the small polaron eigenstates when compared with the 1D Davydov model. However, the origin of these modifications is twofold. First, the vibron dynamics is different because an helix allows for vibron hops between different spines of hydrogen-bonded peptide units. Then, the helix supports both acoustical and optical phonons which participate in a different way to the dressing effect.  

Nevertheless, to discuss and interpret these observed features, let us first mention some general remarks about the modification of the Davydov problem. Indeed, according to the 1D model, the vibron-phonon dynamics along a single spine of hydrogen-bonded peptide units reduces to that of a chain formed by the sites $n$, $n+3$, $n+6$ ... . 
In a similar way, since a real 3D helix is invariant under a rotation of angle $\theta_0$ followed by a translation of length $h$, the vibron-phonon dynamics reduces to that of a 1D lattice formed by all the sites $n$, $n+1$, $n+2$, $n+3$, ... . As the result, in a 3D helix, both the vibron and the phonon reduced wave vectors lye in a Brillouin zone three times greater than the Brillouin zone introduced in the 1D model. In addition, the vibron-phonon dynamics is now governed by long range interactions. Interactions between nearest and second nearest neighbor sites characterize coupling between different spines of hydrogen-bonded peptide units. By contrast, interactions between third nearest neighbor sites account for the coupling in a given spine. These latter couplings correspond to the nearest neighbor interactions in the 1D Davydov model.

The analysis of the collective motion of the residues has revealed that the phonon spectrum exhibits an acoustical branch. In the long-wavelength limit, this branch characterizes the longitudinal sound wave which propagates along the spines of hydrogen-bonded peptide units, as in the 1D Davydov model. However, since the residues are allowed to move in a 3D space, new features occur in a real $\alpha$-helix. First, the spectrum exhibits an optical branch associated to a high frequency breathing motion of the radius of the helix. More precisely, this radial motion takes place in the long-wavelength limit whereas hybridizations with a torsional motion of the helix backbone occur as when increasing the phonon wave vector. In other words, optical phonons originate in the covalent bonds between different spines of hydrogen-bonded peptide units. Then, in a marked contrast with the 1D Davydov model, the polarization of the acoustic branch strongly depends on the phonon wave vector. More precisely, a strong hybridization between longitudinal and radial motion takes place when $k$ reaches $2\pi/3$. In that case, the acoustical branch does not refer to longitudinal sound wave anymore. 

In fact, the wave vector $k=2\pi/3$ plays a key role for acoustical phonons. Near $k=2\pi/3$, the shape of the acoustic branch strongly depends on the geometry of the helix backbone through the value taken by the $\theta_{0}$ angle. Indeed, when $\theta_{0}=2\pi/3$, our calculations have shown that this branch tends to zero (not drawn in Fig. 2). 
This surprising feature can be understood as follows. When $\theta_{0}=k=2\pi/3$, it is straightforward to show that the local frames $n$, $n\pm3$, $n\pm6$, ... are equivalent and that all the motions of the corresponding residues are in phase since $\exp(ikn)=\exp(ik(n\pm3))=\exp(ik(n\pm6))=...$. Therefore, these motions correspond to a zero frequency uniform translation along the axis of the helix so that the physics is equivalent to that of the 1D model at the center of its own Brillouin zone. When $\theta_{0}$ decreases to reach the value characterizing the $\alpha$-helix conformation, i.e. $\theta_0=100^{o}$, the equivalence between the local frames $n$, $n\pm3$, $n\pm6$, ... is broken. As a result, in phase motions of the associated residues do not correspond to a uniform translation anymore but are connected to an hybridization between radial and torsional motions. Therefore, a non vanishing frequency occurs in the spectrum.  

As mentioned previously, the vibron dynamics in a 3D $\alpha$-helix is characterized by two kinds of hopping processes. The first kind refers to vibrational transition between different spines of hydrogen-bonded peptide units whereas the second kind involves vibron hops along the same spine. In that context, the nature of the vibron dynamics essentially results from the competition between these two kinds of processes.  
Within the undressed limit, i.e. when the vibron-phonon interaction is disregarded, vibron transitions between nearest neighbor and third nearest neighbor sites represent the dominant hopping mechanism since $J(1)=-12.4$ cm$^{-1}$ and $J(3)=7.8$ cm$^{-1}$ whereas $J(2)=3.9$ cm$^{-1}$. In other words, the 3D nature of the helix plays a crucial role and the vibron is delocalized between the different spines. However, our calculations have clearly shown that the dressing effect strongly modifies the value of the effective hopping constants and affects in a different way the two kinds of hopping processes. Indeed, at low temperature  $\mid J_{eff}(1)\mid > \mid J_{eff}(3)\mid>\mid J_{eff}(2)\mid $ so that the small polaron behaves almost like the undressed vibron. This is no longer the case as when the temperature is increased because the two kinds of hopping processes do not experience the same dressing effect. Indeed, the dressing for vibrons hops between different spines is more efficient than the dressing which affects vibron transitions along the same spine. As a result, above a critical temperature $\mid J_{eff}(3)\mid$ becomes the dominant hopping constant. In others words, the dressing effect strongly reduces the vibrational exchange between different spines so that the small polaron tends to propagates almost along a single spine of hydrogen-bonded peptide units. Therefore, at biological temperature, the small polaron dynamics is rather similar to the dynamics described by the standard 1D Davydov model. Note that the crossover between the two regimes, i.e. inter-spine and intra-spines vibron hops, is also induced by both an increase of the small polaron binding energy and a modification of the $\theta_0$ angle which defines the helix backbone geometry. In this latter case, when $\theta_0$ tends to $2\pi/3$, the vibrational exchanges between different spines vanish and only $\mid J_{eff}(3)\mid$ remains finite (see Fig. 5c).

These features essentially originate from the coupling between the vibron and the acoustical phonons of the helix. Indeed, our calculations have revealed that the contribution to the dressing of the optical phonons is rather small, i.e. typically about two orders of magnitude smaller than the contribution of the acoustical phonons. This effect comes from the fact that the coupling constant $S_{o}(n,T)$ is proportional to $\mid \chi_{1}/K_{1}\mid^{2}$ whereas $S_{a}(n,T)$ varies according to $\mid \chi_{3}/K_{3}\mid^{2}$. Since $\chi_{1}$ is smaller than $\chi_{3}$ and $K_{1}$ is greater than $K_{3}$, it is straightforward to show that $S_{o}(n,T)<<S_{a}(n,T)$. 

At this step, the role played by the $\theta_{0}$ angle as well as the fact that inter-spine or intra-spine vibron hops are subjected to a different dressing suggest that the observed features originate in the singular behavior of the acoustical phonons near the wave vector $k=2\pi/3$. Indeed, to understand more clearly this effect, let us assume an helix conformation with $\theta_{0}=2\pi/3$. As mentioned previously, in that case, the local frames $n$, $n\pm3$, $n\pm6$, ... are equivalent. When $k=2\pi/3$, the residues $n$, $n\pm3$, $n\pm6$, ... move in phase according to a zero frequency uniform translation. As a consequence, the frequency of the $n$th amide-I vibration does not experience any modulation due to the motion of these residues. By contrast, the coupling constant $S_{a}(n,t)$ with residues located in different spines diverges since the phonon frequency is equal to zero. It results from this behavior around $k=2\pi/3$ that the hopping constant along a given spine is slightly affected by the dressing effect whereas hopping constants between different spines vanish. When $\theta_{0}$ decreases to $100^{o}$, the equivalence between the local frames $n$, $n\pm3$, $n\pm6$, ... is broken. Nevertheless, the motion of the residues along a given spine are more in phase than the motion due to residues located onto different spines. 
As a consequence, the modulation of a particular amide-I vibration due to the motion of residues located on different spines is more important than the modulation produced by residues located in the same spine. Therefore,  
the dressing effect experienced by a vibron moving along a given spine is always smaller than the dressing effect which affects vibron hops between different spines. Note that the previous results corroborate the calculations performed by Hennig \cite{kn:hennig} who have shown, within the soliton approach, that the preferred transport path takes place along a given spine whereas inter-spine energy transfer is suppressed. 
 
Finally, let us mention that the creation of a vibron in a 3D helix is accompanied by a deformation of the helix backbone different than the distortion which occurs in the 1D Davydov model. Indeed, we have shown that the residues located on the spine where the vibron has been excited experience a more intense deformation than the other residues. This feature originates in the fact that inter-spine covalent bonds are more rigid than intra-spine hydrogen bonds. As a consequence, a distortion propagates more easily along a spine of hydrogen-bonded peptide units than along a covalent bond. The main consequence is that residues located on the excited spine do not react as residues located on the two other spines so that an asymmetric deformation of the helix backbone occurs. Due to this asymmetric behavior, the radial displacement of the residues located on the excited spine exhibits an important decrease whereas the radial displacement of the other residues just slightly increases. In the same way, both the longitudinal and the torsional deformations appear as kink-like solitons characterized by oscillations which account for the fact that the distortion essentially affects the excited spine, only. 

In that context, we basically recover the results obtained by Hennig \cite{kn:hennig} although our 3D model slightly differs from its cylinder model. Indeed, as Hennig, we have shown that the longitudinal distortion in a 3D helix is smaller than the distortion occurring in the standard 1D model. This feature is due to the fact that in a 3D helix inter-spine covalent bonds tend to prevent the deformation of the hydrogen bonds. In addition, Hennig has shown that the helix radius decreases in the excited region according to a deformation two orders of magnitude smaller than the longitudinal distortion. As explained previously, the helix model we use does not reduce to a cylinder. Therefore, when a vibron is excited on given site, the three spines react in an asymmetric way. 
Nevertheless, the decrease of the helix radius was obtained, but for the residues belonging to the excited spine, only. In addition, we have shown that the amplitude of the radial deformation is about the same order of magnitude than the longitudinal distortion. In a marked contrast with Hennig who has assumed that the decrease of the radius originates in the optical phonons only, our model suggests that both optical and acoustical phonons strongly affect the radial deformation. In fact, our calculations have clearly shown that both the radial and the longitudinal distortions are essentially governed by the acoustical phonons so that they have the same order of magnitude.

\section{conclusion}

In the present paper, the vibron dynamics associated to amide-I modes in a 3D $\alpha$-helix has been described. The helix was modeled by three spines of hydrogen-bonded peptide units linked via covalent bonds. To remove the strong coupling between the vibrons and the phonons associated to the external motion of the residues, a Lang-Firsov transformation was applied and a mean field procedure was performed to obtain the dressed vibron point of view. It has been shown that the vibron dynamics, due to the 3D nature of the helix, results from the competition between two kinds of hopping processes. The first kind refers to vibrational transitions between different spines whereas the second kind involves vibron hops along the same spine. Our study has revealed that several parameters such as the temperature, the small polaron binding energy and the helix backbone conformation, allow for a transition between two regimes. Indeed, at low temperature or weak small polaron binding energy, the polaron behaves as an undressed vibron delocalized between the different spines. By contrast, at biological temperature or strong small polaron binding energy, the dressing effect strongly reduces the vibrational exchanges between different spines so that the polaron tends to propagate along a single spine, only. The occurrence of these two regimes also depends on the helix backbone geometry via the value of the $\theta_{0}$ angle and it has been shown that when $\theta_0$ tends to $2\pi/3$ inter-spine vibron hops are suppressed. Although the phonon spectrum exhibits both an acoustic branch and an optic branch, it has been shown that the previous features originate in the coupling between the vibrons and the acoustical phonons which exhibit a singular behavior near $k=2\pi/3$. 

Finally, the previous results clearly show that the standard 1D Davydov model is a rather good approximation at biological temperature for which the dressing effect drastically reduces inter-spine vibron hops. In that case, for the vibron dynamics, the helix can thus be viewed as formed by three independent spines. 

To conclude, let us mention that the second paper of this series \cite{kn:helixII} is devoted to the generalization of the present work when two vibrons are excited. In that case, both the intramolecular anharmonicity and the strong vibron-phonon coupling act as nonlinear sources which favor the occurrence of specific states called two-vibron bound states. As shown in paper II, these states exhibit an experimental signature within nonlinear pump-probe spectroscopy and they are expected to play a key role for energy storage in helices due to their formal resemblance with quantum breathers.

\appendix

\section{General expression of the dynamical matrix}

In a general way, the force constant tensor Eq.(\ref{eq:PHI}) is expressed as 
\begin{eqnarray}
\Phi_{\alpha\beta}(nn')&=&-\frac{V'(n n')}{R(n n')}\delta_{\alpha \beta}  \\
&-&\left[ V''(n n')-\frac{V'(n n')}{R(n n')} \right]\frac{R_\alpha(n n')R_\beta(n n')}{R(n n')^2} \nonumber 
\label{eq:PHI2}
\end{eqnarray}
where $V'(n n')$ and $V''(n n')$ denote the first and the second derivative of the pair potential between the residues $n$ and $n'$, respectively. Note that the force constant tensor satisfies the well-known relation 
$\sum_{n'} \Phi_{\alpha\beta}(n n') = 0$.

Within the representation of the displacements of the residues in the set of local frames, the transformed force constant tensor $\bm{\Psi}$ is expressed in terms of the local rotational matrices $\bm{T}(n)$ (Eq.(\ref{eq:T})), as 
\begin{eqnarray}
\bm{\Psi}(nn')&=&-\frac{V'(n n')}{R(n n')}\bm{T}(n-n') \\
&-&\left[ V''(n n')-\frac{V'(n n')}{R(n n')|} \right]\frac{R_0^2}{R(n n')|^2} \bm{A}(n n') \nonumber
\label{eq:PSI}
\end{eqnarray}
where the matrix $\bm{A}(n n')$, which depends on the index difference $n-n'$, is written in terms of the reduced parameter $\tilde{h}=h/R_0$ as

\begin{widetext}
\begin{equation}
\bm{A}(0n) =
\begin{pmatrix}
-(1-\cos{n\theta_0})^2 & -(1-\cos{n\theta_0})\sin{n\theta_0} & -(1-\cos{n\theta_0})n\tilde{h} \\
(1-\cos{n\theta_0})\sin{n\theta_0} & \sin^2{n\theta_0} & \sin{n\theta_0}n\tilde{h} \\
(1-\cos{n\theta_0})n\tilde{h} & \sin{n\theta_0}n\tilde{h} & (n\tilde{h})^2
\end{pmatrix}
\end{equation}
\end{widetext}
At this step, from Eq.(A3) and by using the notations introduced by Christiansen et al. \cite{kn:chris2}, the dynamical matrix $\bm{D}(k)=\sum_{n}\bm{\Psi}(0n)\exp(ikn)/M$ is finally defined as 
\begin{equation}
\bm D (k) =
\begin{pmatrix}
 c_{11} & ic_{12} & ic_{13} \\
-ic_{12} &  c_{22} & c_{23} \\
-ic_{13} &  c_{23} & c_{33}
\end{pmatrix}
\end{equation}
where 
\begin{eqnarray}
c_{11}&=&\sum_n  2\alpha_n[1-\cos(n\theta_0)]^2[1+\cos(nk)] \nonumber \\
&+&2\beta_n[1-\cos(n\theta_0)\cos(nk)]  \nonumber \\
c_{12}&=&\sum_n 2\alpha_n[1-\cos{n\theta_0}]\sin(n\theta_0)\sin(nk) \nonumber \\
&+&2\beta_n\sin(n\theta_0)\sin(nk)  \nonumber \\
c_{13}&=&\sum_n 2\alpha_n nh[1-\cos{n\theta_0}]\sin(nk)  \\
c_{22}&=&\sum_n 2\alpha_n\sin^2(n\theta_0)[1-\cos{nk}] \nonumber \\
&+&2\beta_n[1-\cos(n\theta_0)\cos(nk)]  \nonumber \\
c_{23}&=&\sum_n 2\alpha_n nh\sin(n\theta_0)[1-\cos{nk}] \nonumber \\
c_{33}&=&\sum_n 2\alpha_n(nh)^2[1-\cos{nk}] \nonumber \\
&+&2\beta_n[1-\cos(nk)] \nonumber
\end{eqnarray}
where the summation is performed over $n=1,2,3,...$ and where 
\begin{eqnarray}
\alpha_n &=& \frac{R_0^2}{M R(0n)^2}\left[ V''(0 n)-\frac{V'(0 n)}{R(0 n)} \right] \nonumber \\
\beta_n  &=& \frac{V'(0 n)}{M R(0 n)}
\end{eqnarray}
Note that, to reduce the number of parameters, we have assumed in our work that the helix conformation is stabilized when all the pair potentials are at equilibrium so that $V'(0n)=0$ and $K_{\mid n \mid}=V''(0n)$.

\section{Distortion of the helix backbone}

By disregarding the remaining coupling Hamiltonian $\Delta H$, the lattice distortion onto the $n$th residue is expressed, in the local frame, as
\begin{equation}
\bm{x}_n=\langle\tilde{\psi}|U \bm{v}(n) U^{-1}| \tilde{\psi} \rangle
\label{eq:dist1}
\end{equation}
where $|\tilde{\psi}\rangle= b^\dag_{n_0}|\varnothing \rangle \otimes |ph\rangle $ denotes the state associated to the creation of a polaron on the $n_{0}$th residue from the vacuum state $|\varnothing \rangle$ and where $|ph\rangle$ stands for the phonon wave function. By using Eq.(\ref{eq:vmu}) and after some straightforward calculations this distortion is finally written as
\begin{equation}
x_{\mu n} = -\sum_{k\sigma} \sqrt{\frac{\hbar}{2NM}}  \frac{\Delta_{k\sigma}\epsilon^\mu_{k\sigma}}{\Omega_{k\sigma}^{3/2}}e^{ik(n-n_0)} + c.c. 
\end{equation}
where c.c. denotes the complex conjugate. From Eq.(B2), the z deformation corresponds to $\delta z_{n}=x_{zn}$ whereas 
the radius deformation $\delta R_n$ and the angle deformation $\delta \theta_n$ are defined as  
\begin{eqnarray}
&&\delta R_n=\sqrt{(R_0+x_{rn})^2+x_{\theta n}^2}-R_{0}\\
&&\delta \theta_n = \tan^{-1}(\frac{(R_0+x_{rn})\sin{n\theta_0} + x_{\theta n}\cos{n\theta_0}} {(R_0+x_{rn})\cos{n\theta_0} - x_{\theta n}\sin{n\theta_0}})-n\theta_0 \nonumber
\end{eqnarray}

\end{document}